\documentclass[12pt,onecolumn,aps,pra,showpacs,bibfootnote,preprintnumbers, mathtools,superscriptaddress,amsmath,amssymb]{revtex4-2}

\begin{document}

\title{Lotka--Volterra Model with Mutations \\ and Generative Adversarial Networks}

\author{Sergei V. Kozyrev}
\email{kozyrev@mi-ras.ru}
\affiliation{Steklov Mathematical Institute of Russian Academy of Sciences, Gubkina str. 8, Moscow 119991, Russia}

\begin{abstract}
A model of population genetics of the Lotka--Volterra type with mutations on a statistical manifold is introduced. Mutations in the model are described by diffusion on a statistical manifold with a generator in the form of a Laplace--Beltrami operator with a Fisher--Rao metric, that is, the model combines population genetics and information geometry. This model describes a generalization of the model of machine learning theory, the model of
generative adversarial network (GAN), to the case of populations of generative adversarial networks. The introduced model describes the control of overfitting for generating adversarial networks.
\end{abstract}

\maketitle

\section{Introduction}

This paper will discuss the analogies between population genetics and machine learning theory.
The generative adversarial network (GAN) model \cite{Goodfellow}, \cite{Goodfellow1} describes some hybrid of learning theory (maximum likelihood method) and game theory (minimax problem) model for two players, called discriminator and generator. GANs were found to be highly resistant to overfitting, which can be considered as a regularization of the maximum likelihood method by a game theory model. In this paper, we will construct a model of population genetics of the Lotka--Volterra type with mutations on a statistical manifold, which generalizes the model of GANs in learning theory. In the previous works of the author, the applications of the Eigen's model of population genetics in learning theory \cite{PopGenetics}, \cite{Entropy} have been studied. Analogies between the learning theory and the theory of evolution were discussed by A. Turing \cite{Turing} and attract interest at the present time \cite{VWKK}.

It can be assumed that the efficiency of GANs is of the same nature as the effect of accelerating evolution in the presence of predators known in the theory of evolution. Within the framework of the proposed model for GANs, the predator is the generator, the prey is the discriminator, their interaction is described by some version of Lotka--Volterra model. Mutations are described by diffusion on statistical manifolds (of discriminator and generator, respectively); distributions of discriminators (generators) on the corresponding statistical manifolds are considered as populations of prey (predators) with different genomes; adaptation is the concentration of the discriminator and generator populations in the neighborhood of the solution of the minimax problem for GANs. Thus, the model combines the methods of machine learning theory, population genetics and information geometry.

The overfitting control in the considered model is as follows --- the interaction between the discriminator and the generator is designed so that the learning algorithm suppresses narrow peaks of the population at sharp maxima of the likelihood function, and forces the population to occupy wider regions in the hypothesis space. In accordance with the principle of algorithmic stability (see Section 4), sharp peaks correspond to the overfitting regime. In terms of population genetics, this means that a more specialized predator hunts more efficiently.

The exposition of this work is as follows. Section 2 outlines the GAN model; Section 3 introduces its population generalization in the form of the Lotka--Volterra model with mutations, as well as a variant of the Eigen's model of population genetics within the framework of information geometry; Section 4 discusses the control of overfitting for the considered models of population genetics; Section 5 (the Appendix) provides definitions on the following topics: maximum likelihood method; game theory; Lotka--Volterra model; the Lotka--Volterra model with mutations; information geometry.

\section{Generative Adversarial Networks}

Let us discuss the model of generative adversarial networks (GANs) \cite{Goodfellow}, \cite{Goodfellow1}, see also \cite{Nikolenko}.
Let $X$ be the space of objects under investigation; $p_{\rm data}$ is the training sample (distribution) on $X$. The GAN model uses two parametric probability distributions:
discriminator $D(x, \theta_d)$ with parameter $\theta_d$; generator $p_{\rm gen}(x,\theta_g)$ with parameter $\theta_g$.

The GAN model includes the following optimization problems: the discriminator maximizes the proximity to the data $p_{\rm data}$ and the distance from the generator $p_{\rm gen}$
$$
V(\theta_d,\theta_g)=E_{x\sim p_{\rm data}}\log D(x,\theta_d) + E_{x\sim p_{\rm gen}}\log(1-D(x,\theta_d))=
$$
\begin{equation}\label{V}
=\int_{X}p_{\rm data}(x)\log D(x,\theta_d) dx + \int_{X} p_{\rm gen}(x,\theta_g) \log(1-D(x,\theta_d)) dx.
\end{equation}

In fact, in \cite{Goodfellow}, \cite{Goodfellow1} a slightly different (equivalent) construction was used: instead of a parametric family $p_{\rm gen}$ of generator distributions on $X$, a distribution on some auxiliary space and a parametric family of mappings of this space in $X$ were used.

In this case, the data distribution is a (training) sample $\{x_i\}$, $i=1,\dots,n$, then the above functional takes the form of an empirical sum
\begin{equation}\label{V_emp}
V_{\rm emp}(\theta_d,\theta_g)=\frac{1}{n}\sum_{i=1}^n \log D(x_i,\theta_d)  + \int_{X} p_{\rm gen}(x,\theta_g) \log(1-D(x,\theta_d)) dx.
\end{equation}

The generator should minimize the value (in $\theta_g$)
$$
E_{x\sim p_{\rm gen}}\log(1-D(x,\theta_d))=\int_{X} p_{\rm gen}(x,\theta_g) \log(1-D(x,\theta_d)) dx.
$$

The logarithm argument must be positive, which sets a constraint on the parametric family of discriminators --- $D(x,\theta_d)$ must be less than one.

Combination of these two problems gives the following game theory problem (minimax problem, Nash equilibrium): find $\theta_d$ and $\theta_g$ which give the minimax
$$
\min_{\theta_g} \max_{\theta_d} V_{\rm emp}(\theta_d,\theta_g).
$$

In this case, the distribution of the generator $p_{\rm gen}(x,\theta_g)$ (where $\theta_g$ is the solution to the minimax problem) is the solution to the learning problem for the GAN model.

The second term in (\ref{V_emp}) (addition to the empirical sum) and the minimax problem help to control overfitting in the maximum likelihood estimate problem (maximization of the first term in (\ref{V_emp})).

\medskip

\noindent{\bf Evolutionary game theory}. The game theory model considered above has the following biological interpretation. The training sample $p_{\rm data}$ is the grass distribution, the discriminator $D$ is the herbivore (prey) distribution, and the generator $p_{\rm gen}$ is the carnivore (predator) distribution. Optimization problems are interpreted as evolution problems: prey $D$ is looking for grass and running away from predators, predator $p_{\rm gen}$ is looking for prey. Parameters $\theta_d$ and $\theta_g$ (parameters for discriminator and generator) are genotypes of the prey and predators, the variation of these parameters means the presence of genotype mutations; optimization problem means Darwinian evolution; adaptation is the convergence of the discriminator and generator to the solution of the minimax problem for the generative adversarial network.

It is known that the appearance of predators causes an acceleration of evolution. We discuss this effect from the point of view of learning theory. For learning theory, this looks like adding a zero-sum contribution to the empirical risk, and stimulating evolution by predators is a regularization of the learning problem to control overfitting.

\section{Population genetics generalization of GANs}

The transition from the consideration of individual systems to the consideration of ensembles of systems in physics corresponds to the transition from mechanics to statistical mechanics; in the theory of biological evolution, to the transition from Darwinism to population genetics. It seems that such an approach is also fruitful in the theory of machine learning. In the standard approach, the learning algorithm performs the optimization of some functional on the hypothesis space; in the population approach, such an algorithm transforms a population of learning systems on the hypothesis space to a population concentrated in the vicinity of the optimum of the learning problem. In the population approach to learning, some aspects of learning theory become more understandable, in particular, overfitting (see the next section).

Let us construct a model of the Lotka--Volterra type with mutations, which describes a generalization of the GAN model for the case of populations (distributions of networks by parameters).
Namely, we will consider a Lotka-Volterra model, where the discriminator is the prey and the generator is the predator. Populations of discriminators and generators live on their statistical manifolds; in this model, the statistical manifold is the space of genomes, and diffusion on this space describes mutations. Learning for this model will be given by the convergence of the discriminator and generator populations (for the model of GANs, mainly the generator population) to populations concentrated in the neighborhood of the solution of the learning problem; i.e. a hypothesis for the generator giving a peak in the population of the generator can be considered a solution for the learning problem.

Let us introduce the following notations: let $X$ with parameter $x$ be the data space;
$Y$ with parameter $y$ is the space of (genomes) of the discriminator;
$Z$ with parameter $z$ is the space (of genomes) of the generator.

Thus, in addition to the distribution of data on $X$, we consider populations of discriminators $f(y,t)$ on $Y$ and generators $g(z,t)$ on $Z$ (not necessarily normalized).

Let the dynamics of the discriminator population be described by the diffusion equation on the statistical manifold $Y$, the diffusion generator is the Laplace--Beltrami operator for the Fisher metric (see the Appendix)
\begin{equation}\label{ddtf}
\frac{\partial}{\partial t} f(y,t)=M_d\Delta_{y}f(y,t)+ A(y) f(y,t)- N_d f(y,t)\int_{Z} B(y,z)g(z,t)dz;
\end{equation}
similarly for the population dynamics of the generator on $Z$
\begin{equation}\label{ddtg}
\frac{\partial}{\partial t} g(z,t)=M_g\Delta_{z}g(z,t) - C g(z,t) + N_g g(z,t)\int_{Y} B(y,z)f(y,t)dy;
\end{equation}
terms with functions $A$ and $B$ are taken from the minimax problem for (\ref{V}), (\ref{V_emp}), $A(y)$ describes feeding the discriminator on the data, or fitness of the discriminator (the likelihood, if $p_{\rm data}$ is a training set)
\begin{equation}\label{Ay}
A(y)=\exp\left(\int_{X}p_{\rm data}(x)\log D_y(x) dx\right)=\prod_{i=1}^n  D_y(x_i);
\end{equation}
$B(y,z)$ (where $B(y,z)\ge 0$) describes predation of a generator on a discriminator
\begin{equation}\label{Byz}
B(y,z)= -\alpha\int_{X} p_{{\rm gen},z}(x) \log(1-D_{y}(x)) dx;
\end{equation}
$\alpha>0$, the term with $C>0$ for the generator describes the extinction of the predator in the absence of prey, the constants $M_d,M_g>0$ regulate the rate of mutations for the discriminator and generator, respectively, $N_d,N_g>0$ regulate the interaction of the discriminator and generator.

Here, for the term with function $A$  compared to (\ref{V}), (\ref{V_emp}), an exponent is applied to get a positive function (the logarithm can be non-positive). The $A$ function also depends on the data (on the training sample), this function is concentrated on the values of the parameter (hypothesis) $y$ of the discriminator for which the discriminator is close enough to the distribution of the data. The training sample for $A(y)$ in population genetics models may depend on time-varying external conditions, that is, contain time-dependent noise (in this case $A(y)$ depends on time).

The values of the function $B(y,z)$ of the form (\ref{Byz}) decrease as the generator distribution moves away from the discriminator distribution, i.e. wide peaks of the discriminator are eaten away more slowly, or a more specialized predator hunts more efficiently. This implies for the (\ref{ddtf}), (\ref{ddtg}) system the effect of suppressing narrow population peaks for the discriminator and generator --- the corresponding populations will tend to occupy wider regions on their statistical manifolds; this is also facilitated by the presence of diffusion in the equations (\ref{ddtf}), (\ref{ddtg}), but an additional effect of narrow peak suppression will also be given by nonlinear terms.

In fact, for the described effect, the specific form (\ref{Byz}) (as in the work \cite{Goodfellow}) for the function $B(y,z)$ is not important, it is important that this function decreases as the discriminator moves away from the generator. Therefore, we can choose this function in the ''Coulomb'' form
\begin{equation}\label{Byz1}
B(y,z)=D(p_{{\rm gen},z}|D_{y})^{-1}=\left(\int_{X}p_{{\rm gen},z}(x)\log\frac{p_{{\rm gen},z}(x)}{D_{y}(x)}dx\right)^{-1},
\end{equation}
where $D(p|q)$ is the Kullback--Leibler distance.
For GAN, the generator is attracted to the discriminator, and the discriminator is repelled from it (the ''Coulomb'' interaction is asymmetric).

It should be noted that the Nash equilibrium (solution of the minimax problem) need not necessarily exist in the class of pure strategies. From the point of view of the Lotka--Volterra type model, this is related to the possibility of the existence of population cycles; for a model with mutations, such cycles can include not only population oscillations, but also changes in genomes.

Based on the equation (\ref{ddtf}), we propose an analogue of the Eigen's model \cite{Eigen} for the information geometry, namely the equation
\begin{equation}\label{Eigen}
\frac{\partial}{\partial t} f(y,t)=M_d\Delta_{y}f(y,t)+ A(y) f(y,t)- \frac{1}{N} f(y,t)\int_{Y} \left(M_d\Delta_{u}+A(u)\right)f(u,t)du;
\end{equation}
see also \cite{PopGenetics}, \cite{Entropy} for a discussion of the application of Eigen's model of population genetics to learning theory.

In such a model, the total population (the integral $\int_Y f(y)dy$) will be conserved if it is equal to $N$, because the multiplier in the last term does not depend on $y$. For the equation of the Lotka--Volterra type, this is no longer the case.

In the present model, information geometry is used to introduce the Fisher metric and the Laplace--Beltrami operator on the statistical manifold. For an overview of information geometry in the classical and quantum case, see \cite{Chentsov}, \cite{Chentsov1}, \cite{Amari},\cite{Amari1}, \cite{Gibilisco}, categorical aspects of information geometry are discussed in \cite{ Manin}.

\section{Control of overfitting}

Let us discuss different regimes in learning theory (using the example of maximum likelihood estimation).

Underfitting is a situation where we are unable to obtain a maximum likelihood estimate (we cannot solve an optimization problem).

The desired outcome of learning is the reproducibility of learning on the control sample (generalization ability), that is, the maximum likelihood estimate should be weakly dependent on the sample for the majority (in probability) of samples (that is, for generic samples).

Overfitting is a situation in learning theory where there is a high likelihood (low risk) on the training sample, but a low likelihood on the control sample, that is, there is a strong dependence of the  maximum likelihood estimate on the sample. The effect is related to the fitting of the learning model hypothesis to the training sample (including its random details), as a result, the maximum likelihood is achieved on hypotheses that are far from the correct one, which manifests itself in low likelihood on the control sample (low generalization ability).

The following approaches were proposed to control overfitting:
The learning model should have a finite and small VC-dimension (the learning model should be simple) \cite{Vapnik};
Risk functional minima should be flat \cite{FlatMinima} (which is in this paper was related to the principle of minimum description length for the learning model);
Uniform, algorithmic and CV stability \cite{Stability}, \cite{AlgStability}, \cite{CV-loo} of the solution of the learning problem (variants of stability of the solution of the learning algorithm to perturbations of the training sample) were discussed.

From this point of view, if the solution to the learning problem has the ability to generalize, then this solution corresponds to the maximum for a relatively wide likelihood peak in the hypothesis space.
Narrow likelihood peaks should correspond to overfitting. In addition, such peaks should be far from solutions of the learning problem which correspond to broad peaks.
Therefore, \emph{learning is finding the maximum of a sufficiently wide likelihood peak in the hypothesis space for a generic training sample}.

\medskip

The control of overfitting in terms of the Eigen--type model (\ref{Eigen}) is as follows: for narrow population peaks on overfitted hypotheses, mutations to nearby hypotheses (where fitness is low) lead to extinction of the population. The possible dependence of the function $A(y)$ of the form (\ref{Ay}) on time (through the dependence of the data on time) enhances this effect --- the fitness peak can shift relative to the population peak, which reduces fitness for narrow peaks (this is exactly the effect of the lack of algorithmic stability when the sample is modified). A similar mechanism was discussed by R. Fisher \cite{Fisher}, that is, Fisher actually discussed some version of algorithmic stability for population genetics.
On the other hand, too high mutation rate leads to smearing of all peaks, including those corresponding to the correct hypothesis (the peak of fitness for which should be wide enough). Too high mutation rate within the framework of Eigen's model is discussed as a catastrophe of errors and is expressed in a decreasing in the efficiency of purifying selection in population genetics.

\medskip

Control of overfitting from the point of view of the Lotka--Volterra type model with mutations (\ref{ddtf}), (\ref{ddtg}) has the following form: the matching the hypothesis of a predator (generator) and a narrow peak of the population of a prey (discriminator) leads to rapid reproduction of the predator and eating away the sharp peak. This forces the discriminator to occupy wider peaks in the hypothesis space, since the values of the $B(y,z)$ function of the form (\ref{Byz}) or (\ref{Byz1}) decrease as the generator distribution moves away from the discriminator distribution, that is, wide peaks of the discriminator are eaten away more slowly, the more specialized predator hunts more efficiently. This leads to a similar behavior for the generator, with these wider peaks being concentrated on hypotheses with the presence of data (the training sample).
This should lead to reduced overfitting by suppressing the narrow peaks of population of the discriminator and generator. The other two control mechanisms for overfitting are the same as for the Eigen's-type model described above -- mutations and the possible dependence of $A(y)$ on time. The rate of predator reproduction is higher than the rate of mutations, therefore the Lotka--Volterra type model with mutations provides better overfitting control than the Eigen's type model; it also explains the acceleration of biological evolution in the presence of predators.

\medskip

Thus, population genetics \cite{Fisher} and evolutionary game theory \cite{MaynardSmith} models seem to be directly related to modern machine learning models. The introduced Lotka--Volterra type model with mutations (\ref{ddtf}), (\ref{ddtg}) can be considered as a population generalization of the GAN model (the Lotka--Volterra type model with mutations for GAN, or LVM--GAN). Such a model generates solutions of the learning problem that are resistant to overfitting in the sense of \cite{Stability}, \cite{CV-loo} (corresponding to flat \cite{FlatMinima} likelihood maxima). The Darwinian adaptation for this model of population genetics is the convergence of the discriminator and generator populations to some distributions concentrated in the vicinity of the solution of the minimax problem for GAN. A generalization of the Eigen model (\ref{Eigen}) for the case of information geometry was also introduced.

\section{Appendices}

\noindent{\bf Maximum likelihood method}. Let us consider a parametric family of probability distributions $p(x,\theta)$, $x\in X$ with parameter $\theta$. Let $\{x_i\}$ be a sample of independent trials in $X$ (i.e., we assume the existence of a probability distribution $q$ on $X$ that generates such a sample). The likelihood function has the form of the product of probability densities for events from the sample
$$
L(\{x_i\},\theta)=\prod_{i=1}^n  p(x_i,\theta).
$$
Maximum likelihood estimate $\theta$
$$
\theta_0=\arg \max L(\{x_i\},\theta)=\arg \max \frac{1}{n} \log L(\{x_i\},\theta)=\arg \max \frac{1}{n} \sum_{i=1}^n \log p(x_i,\theta)
$$
takes the form of an empirical sum, we assume that the maximum likelihood distribution $p(x,\theta_0)$ approximates the unknown distribution $q(x)$.

Overfitting is a well-known problem in learning theory: fitting $p(x,\theta)$ to the training sample $\{x_i\}$ can have a high likelihood but a low likelihood on the control sample $\{x'_i\}$ (some other sample generated by the distribution $q$).
The control of overfitting in learning theory usually comes down to regularization (additions to the empirical sum that change the form of the likelihood function, or another kind of empirical risk functional).

\medskip

\noindent{\bf Game theory}.
The game is a function of several variables called players, the game takes values equal to a set of real numbers, one for each player
$$
(v_1,\dots,v_n)(s_1,\dots,s_n),
$$
the $i$-th number $v_i$ in the set is called the payoff for the $i$-th player.
The variable $s_i$ for each player takes on values called (pure) strategies, and each player has its own set of strategies.

Example: two players, zero-sum game (the sum of the players payoffs is zero).

Mixed strategy for the $i$-th player: the probability distribution $p_i(s_j^i)$ for this player's strategies, the index $j$ enumerates different strategies for the $i$-th player. The payoff of the $i$-th player for a set of mixed strategies
$$
\langle v_i\rangle=\sum_{j_1,\dots,j_n} p_1(s_{j_1}^1)\dots p_n(s_{j_n}^n) v_i(s_{j_1}^1,\dots,s_{j_n}^n)
$$
is the average of the payoff of the $i$-th player over mixed strategies for all players.
Here the index $j_k$ runs through the values of possible strategies $s_k$ for the $k$-th player:
$s_k\in\{s_{1}^k,\dots,s_{m_k}^k\}$.

Nash equilibrium: A set of strategies in which no participant can increase the payoff by changing his strategy if the other participants do not change their strategies. Exists in the class of mixed strategies (may not exist in the class of pure strategies).

Maximin: the largest payoff that a given player can get without knowing the actions of other players
$$
\underline{v_i}=\max_{s_i}\min_{s_{-i}}v_i(s_i,s_{-i}),
$$
where $s_i$ is the strategy of the $i$-th player; $s_{-i}$ -- other players' strategies; $v_i$ -- payoff of $i$-th player.

Minimax: The smallest payoff that other players can force without knowing the player's action
$$
\overline{v_i}=\min_{s_{-i}}\max_{s_i}v_i(s_i,s_{-i}).
$$

Minimax is not less than maximin
$$
\overline{v_i}\ge \underline{v_i};
$$

The minimax for two-player zero-sum games is the same as the Nash equilibrium.

\medskip

\noindent{\bf Lotka--Volterra Model} describes the population dynamics of two species (predator and prey) by a system of two ODEs
$$
\frac{dx}{dt}=\alpha x-\beta xy;
$$
$$
\frac{dy}{dt}=-\gamma y+\delta xy;
$$
where $x$ is the prey population; $y$ is the predator population.
All constants are positive; the non-linear term describes the interaction between predator and prey.

For an ecological niche of finite volume, the first equation changes as ($N>0$)
$$
\frac{dx}{dt}=\alpha (x-x^2/N)-\beta xy.
$$

The Lotka--Volterra model with mutations generalizes the above model as follows:
there are prey $x_i$, predators $y_m$, the equations of population dynamics have the form
$$
\frac{dx_i}{dt}=\sum_j A_{ij} x_j -  x_i \sum_{n}B_{in} y_n;
$$
$$
\frac{dy_m}{dt}=\sum_{n} C_{mn} y_n  + y_m \sum_{j} B_{jm} x_j .
$$

The matrix $A$ has the following meaning: the diagonal part describes the reproduction of prey, the off-diagonal part describes mutations (the matrix elements are positive).
Matrix $C$: the diagonal part (with negative matrix elements) describes the extinction of predators, the off-diagonal part describes mutations (these matrix elements are positive).
The $B$ matrix describes the interaction between predator and prey (the matrix elements are positive).

For a limited ecological niche, the first equation takes the form
$$
\frac{dx_i}{dt}=\sum_j A_{ij} x_j - \frac{1}{N} x_i \sum_{ij} A_{ij} x_j -  x_i \sum_{n}B_{in} y_n.
$$
In the absence of predators $y_n=0$, this equation takes the form of the equation of the Eigen's model \cite{Eigen}, which describes the competition of genotypes in a limited ecological niche in presence of mutations.

\medskip

\noindent{\bf Information geometry} \cite{Chentsov}, \cite{Chentsov1}, \cite{Amari}, \cite{Amari1}, \cite{Gibilisco}, \cite{Manin}. A statistical manifold is a manifold of parameters of a parametric probability distribution, or a manifold whose points are probability distributions on $X$ (smooth dependence of the distribution on parameters is assumed).

The Kullback--Leibler distance (non-symmetric) between two probability distributions on $X$ is defined as:
$$
D(p|q)=\int_{X}p(x)\log\frac{p(x)}{q(x)}dx.
$$

The Fisher--Rao metric on the statistical manifold has the form
$$
g_{ij}(\theta)=\int_{X}\frac{\partial \log p(x,\theta)}{\partial \theta_i}\frac{\partial \log p(x,\theta)}{\partial \theta_j} p(x,\theta)dx.
$$

Kullback--Leibler distance expansion for small parameter differences $\Delta\theta=\theta-\theta_0$ is related to the Fisher metric:
$$
D(p(\theta_0)|p(\theta))=\frac{1}{2}\sum_{ij}g_{ij}(\theta_0)\Delta\theta^i\Delta\theta^j.
$$

The Laplace--Beltrami operator on a statistical manifold with parameter $\theta$ has the following form, where $g_{ij}$ is the Fisher metric
$$
\Delta_{\theta}= \frac{1}{\sqrt{g}}\sum_{i}\frac{\partial}{\partial \theta_i}\left(\sqrt{g}\sum_{j}g^{ij}\frac{\partial}{\partial \theta_j}\right).
$$

\bigskip

{\bf Acknowledgments}.

This work is supported by the Russian Science Foundation under grant 19--11--00320, \\
https://rscf.ru/en/project/19-11-00320/.

\end{document}